\documentstyle[amssymb,aps,prb,twocolumn,epsf]{revtex}
\begin{document}
\draft
\title{Pressure Effects and Large Polarons in Layered MgB$\boldmath{_2}$ Superconductor}
\author{V.~A.~Ivanov${}^{a)}$\cite{a1}, M.~A.~Smondyrev${}^{b)}$, J.~T.~Devreese${}^{a)}$\cite{a2}}
\address{${}^{a)}$ Departement Natuurkunde, Universiteit Antwerpen (UIA), \\
Universiteitsplein 1, B-2610 Antwerpen, Belgium}
\address{${}^{b)}$ Bogoliubov Laboratory of Theoretical Physics, \\
Joint Institute for Nuclear Research\\
141980 Dubna, Moscow Region, Russia}
\date{\today}
\maketitle

\begin{abstract}
We consider the dependence of the MgB$_2$
superconducting critical temperature on the pressure.
Our model exploits the influence of the large polarons on
the band structure of the layered MgB$_2$ superconductor. Namely,
the hole Pekar-Fr\"{o}hlich polarons form quasi two-dimensional
potential wells in the boron plane which shift the positions of the
$\sigma$- and $\pi$-bands. This energy shift depends on the pressure and
the Cooper pairing of the correlated
$\sigma$-electrons happens inside polaron wells.
The results obtained are as follows:
$dT_c/dp \simeq -\alpha (5.2\pm 0.9)~K/GPa$
or $dT_c/dp \simeq -\alpha (6.9\pm 1.1)~K/GPa$ for a different choice of the
Gr\"{u}neisen parameter. Being compared with known experimental
data they give us a resonable interval for the value of the
Fr\"{o}hlich electron-phonon coupling constant:
$\alpha \simeq 0.15 - 0.45$.
\end{abstract}


\narrowtext

\section{Introduction}

Soon after the discovery of the superconductivity in MgB$_2$
the pressure influence on the critical temperature $T_c$ has
been experimentally studied. The baric derivatives $dT_c/dp$
were extracted from the obtained data. At the
low-pressure ($p \leq 2\,GPa$) the experimental results are as follows:
$dT_{c}/dp = -1.2~K/GPa$ in Refs.~\onlinecite{lorenz,tissen}
and $-2~K/GPa$ in
Refs.~\onlinecite{lorenz,prassides,saito}.
Experiments at high-pressure lead at the same sign
of the baric derivatives but demonstrate lower values. Namely,
$dT_c/dp = (-0.8, -1.07, -1.11) K/GPa$ in
Refs.~\onlinecite{monteverde,lorenz,tomita}, respectively.
Many conventional low-$T_{c}$ superconductors also reveal the same sign,
$dT_{c}/dp < 0$, but their absolute values are much lower.

Borons in the crystal MgB$_{2}$
are packed in honeycomb layers alternating with hexagonal layers of
magnesium ions. The in-plane distance, $b=1.78~\AA$, between the boron cations
is less than those between magnesium anions, $a= 3.08~\AA$.
The space separating the boron planes is of the size $c=3.52~\AA$.
The magnesium ions are positioned above the centers of hexagons formed by
boron sites and donate their electrons to the boron planes:
Mg$^{2+}\left[{\rm B}^{-}\left(p^{2}\right) \right] _{2}$.
These $p$-electrons form $\sigma$- and $\pi$-bands
and the charged Mg-layers lift the three-fold
degeneracy between $\sigma$- and $\pi$-electrons and shift the electronic
energy bands so that $\pi$-band becomes lower than the $\sigma$-band.
It occurs that the latter crosses the Fermi-level providing
the light and heavy hole formation.
The electronic structure is formed by the narrow energy-band of the two-fold
degenerate $\sigma$-electrons and the wide-band of the $\pi$-electrons.

As to the phonon modes in MgB$_{2}$ they have a sharp cut-off at
about $100~meV$. Below this energy it is supposed the existence of different
phonon modes\cite{yildirim}. Note that the phonon energies
mentioned in Ref.~\onlinecite{yildirim}
could be overestimated because the approximation used
to derive them is valid only for cubic lattices\cite{hase,gaspari}.
The anharmonic in-plane breathing mode $E_{2g}\ (74.5~meV)$
strongly interacts  with the narrow band $\sigma$-electrons\cite{bohnen}.
Due to this fact its energy is lower than that of
the out-of-plane tilting boron mode $B_{1g}\ (87.1~meV)$.
Besides, there are low-energy acoustic modes and
indications are also found\cite{lampakis,muranaka,jorgensen1}
of the existence of phonon modes with extremely low energies.

The most of the theoretical speculations on
the $T_c$ pressure dependence
in MgB$_2$ are based on the semi-phenomenological approaches and
do not involve the microscopic arguments.
In the scope of the conventional BCS theory
$dT_{c}/dp$ is governed by the competition
between a decrease of the density of electronic states at the Fermi level
and an increase of the Debye frequency in the BCS formula for $T_{c}$.
The weak electron-phonon coupling constant\cite{marsiglio} in MgB$_2$
and the absence of the peak in the electronic density of states
at the vicinity of the Fermi-energy\cite{ravindran} cannot provide
the BCS scenario.

On the other hand, the spectroscopic measurements of the superconducting
gap\cite{chen,tsuda,szabo}, the specific heat behavior\cite{junod,junod2},
the low isotope effect\cite{budko,hinks}
(under the substitution of both B and Mg isotopes)
and the pressure effects\cite{schilling} evidence in favor
of the complex mechanism of the superconductivity in MgB$_2$.
The calculated spectral function\cite{knigavko} and
the analysis\cite{marsiglio}
of the reflectance measurements\cite{tu}
show the possibilities of different superconducting mechanisms
beyond the conventional electron-phonon BCS pairing.

The positive baric derivative of $T_{c}$ was
found\cite{hirsch1,hirsch2} in the microscopic model of the hole dressed
superconductivity what opposes the experimental data.

Among optical modes in MgB$_2$ the $E_{2g}$ is the only Raman active mode.
The high-pressure (up to $15~GPa$) Raman experiment\cite{goncharov}
has revealed a large pressure shift of the $E_{2g}$ phonon mode.
The $E_{2g}$ mode does not manifest any softening
at $T_{c}$ (see Ref.~\onlinecite{martinho}) therefore does not provide
the superconducting electron pairing. Nevertheless, it is
responsible for the $T_c$ behavior with pressure as we will see later.
It should be pointed
out that baric experiments have been completed both for pressed powder
and single crystals.  No structural transitions were found in
experiments\cite{prassides,goncharov,bordet} up to the pressure
$40~GPa$.

The goal of the present paper is to calculate the baric derivative
$dT_c/dp$. We suppose that the large polarons are formed in the
boron-planes due to the $\sigma$-electrons interaction with
the optical $E_{2g}$-phonons\cite{yildirim}.
Because of the polaron formation the
$\sigma$-band becomes lower while the $\pi$-electrons are not influenced.
As the events take place in the layered medium we consider
the quasi-two-dimensional polaron. The polaron anisotropy parameter and,
correspondingly, the $\sigma$-band energy shift depend on the
lattice constants influenced by the pressure. Roughly speaking,
the Cooper pairs of the correlated $\sigma$-electrons in
the MgB$_2$ superconducting condensate are plunged into
the pancake polaron. To treat the Cooper paring
we accept the kinematic mechanism\cite{ivanov,ivanov1} in MgB$_2$.

\section{Formulation of the model}

In order to describe the superconducting state of MgB$_2$, we adopt the
model with the strongly interacting $\sigma$-electrons (with a narrow
bandwidth $2w_{1}$) overlapping with the noncorrelated $\pi$-electrons
(with a wide bandwidth $2w_{2}$) of boron ions. The important parameter here
is the distance $r$\ between the centers of $\sigma$- and $\pi$-bands
influenced by the pressure.

The positive Hall current and the thermoelectric power
measurements\cite{lorenz2} indicate the hole conductivity in MgB$_2$.
The polaron radius is given by the expression
\begin{eqnarray}
R \sim \sqrt{\frac{\hbar}{m\omega}} = \frac{87.3~\AA}{\sqrt{(m/m_e)\cdot
(\hbar\omega/1~meV)}},
\label{radius}\end{eqnarray}
where $m$ and $m_e$ are the hole and the vacuum electron masses,
respectively, and $\hbar\omega$ is the low-frequency optical phonon
energy responsible for the polaron formation. Inserting the value
$\hbar\omega \simeq 74.5~meV$ for the $E_{2g}$-phonons
and the light hole band mass\cite{mazin} $m=0.25m_e$
into Eq.~(\ref{radius}) the polaron radius can be estimated
as $R \simeq 20~\AA$ which is much larger than the lattice constant.

Studying the polaron sector of the system we apply the conventional
Pekar-Fr\"{o}hlich Hamiltonian generalized for the consideration of the
peculiarities of MgB$_2$. To take into account the fact that the
$\sigma$-electrons are located in narrow layers inhabited by boron atoms
we introduce the in-plane mass $m$ which is the light hole band mass and the
large mass $M$ in the orthogonal ($z$) direction. The ratio $m/M$
of these masses is related to the probability of a carrier to tunnel from
one layer to another.

To take into account the different polarizability of the medium in the
$xy$-plane and in the $z$-direction we suggest to simulate it by using the
anisotropic Columb potential for the interaction of the hole with the
induced polarization field. That is, we use the
potential of the form $V(\vec r) \propto 1/\sqrt{x^2+y^2+(z/\gamma)^2}$ with
$\gamma \leq 1$ being a phenomenological parameter to describe the
deformation of the Coulomb potential along the $z$-axis (that is, the
oblation of the equipotential surface). In the momentum space the potential
is of the form $\tilde{V}(\vec k) \propto 1/(k_{\parallel }^{2}+\gamma^2
k_{z}^{2})$.

In a sense we study the electronic sector as how all events happen to be
located inside the large polarons. The large polaron being formed we obtain
the lowering of the energy of carriers, that is the renormalization of the
parameter $r$ which becomes dependent on the applied pressure. This
influences drastically the electronic structure of MgB$_2$-like systems,
especially the relative position of $\sigma$- and $\pi$-bands.

\subsection{Polaronic Sector}

Thus, we consider an anisotropic polaron whose motion is confined to
the $xy$-plane while the motion in the perpendicular $z$-direction
(along the crystallographic $c$-axis of MgB$_2$) is restricted.
The Pekar-Fr\"{o}hlich Hamiltonian reads as follows:

\begin{eqnarray}
H &=& {\frac{\vec{p}_{\parallel }^{\,2} }{2m}}+{\frac{p_{z}^{\,2} }{2M}}+
\hbar \omega \sum\limits_{\vec{k}} a_{\vec{k}}^{\dag }a_{\vec{k}}
\nonumber \\
&& +{\frac{1}{\sqrt{\Omega}}}\sum_{\vec{k}}\, \left( a_{\vec{k}}\,V_{\vec{k}}
\,e^{i\vec{k}\cdot \vec{r}}+
a_{\vec{k}}^{\dag }\,V_{\vec{k}}^{\ast }
\,e^{-i\vec{k}\cdot \vec{r}}\right) ,
\label{eq01}
\end{eqnarray}
where $\Omega$ is the crystal volume, $\vec{p}_{\parallel },p_{z}$ are the
hole momenta in the $xy$-plane and in the orthogonal $z$-direction,
respectively,
and $\omega $ is the LO-phonons frequency. Operators $a_{\vec{k}}^{\dag }(a_{%
\vec{k}})\,$ are the creation (annihilation) operators of the phonons with a
momentum $\vec{k}$. The electron-phonon interaction term is specified
by $V_{\vec{k}}$; its squared modulus is the Fourier transform of the Coulomb
potential $V(\vec{r})$. In contrast with the conventional definition we
introduce an additional anisotropy of the Coulomb potential to describe the
different polarizability of the medium in various directions (in the
isotropic boron plane and perpendicular to it):

\begin{equation}
V_{\vec{k}}=-i\hbar \omega \left( {\frac{4\pi \alpha }{ k_{\parallel
}^{2}+\gamma^2 k_{z}^{2}}} \sqrt{\frac{\hbar } {2m\omega }}\right) ^{1/2}.
\label{eq02}
\end{equation}
Here $\alpha$ is the conventional Fr\"{o}hlich coupling constant
of the electron-phonon interaction.

The MgB$_2$ baric compression is anisotropic\cite{tissen,goncharov,vogt}%
. According to Ref.~\onlinecite{goncharov} the compressibility along $c$%
-axis almost twice larger the plane compressibility. Under the hydrostatic
pressure the initial compression along the $c$\ axis is larger than
along the boron plane\cite{jorgensen}. We estimate the parameter $\gamma $
as the ratio of the standard lattice constants: $\gamma \sim c/2a$
(here $c/2$ is the distance between charged boron and magnesium planes).
When the pressure increases the distance between the Mg and the B planes
decreases which is described by the decreasing of the Coulomb potential
anisotropy $\gamma $ in our model. The polaron self-energy $\Delta E$ can be
found within the second order of the perturbation theory:
\begin{equation}
\Delta E=-{\frac{\alpha \hbar \omega }{2\pi ^{2}}}\int {\frac{d^{3}k}{%
(k_{\parallel }^{2}+\gamma ^{2}k_{z}^{2})(k_{\parallel }^{2}+{\frac{m}{M}}%
k_{z}^{2}+1)}}.  \label{eq04}
\end{equation}
The integration in Eq.~(\ref{eq04}) being performed we arrive at the
expression:
\begin{equation}
\Delta E=-\alpha \hbar \omega \sqrt{\frac{M}{m}}\left\{
\begin{array}{ll}
\displaystyle{\frac{1}{\sqrt{1-\Gamma ^{2}}}}\ln {\frac{1+\sqrt{1-\Gamma ^{2}%
}}{\Gamma }}, & \Gamma \leq 1; \\
&  \\
\displaystyle{\frac{1}{\sqrt{\Gamma ^{2}-1}}}{\rm arctan}\sqrt{\Gamma ^{2}-1}%
, & \Gamma \geq 1.
\end{array}
\right.   \label{eq05}
\end{equation}
Here the parameter $\Gamma =\gamma \sqrt{M/m}$. This formula reproduces two
well-known limiting cases. When $M=m$ and $\gamma =1$, then $\Gamma =1$ and
we obtain $\Delta E=-\alpha \hbar \omega $, that is the conventional result
for the three-dimensional polaron. When $\gamma =1$ and $M\gg m$, we arrive
at the result $\Delta E=-{\frac{\pi }{2}}\alpha \hbar \omega $ which is
valid for the two-dimensional polaron confined to a plane\cite{sak,evans}.
If $\gamma $ is finite and $M\gg m$, then $\Gamma \gg 1$ and Eq.~(\ref{eq05}%
) leads at the following expression:
\begin{equation}
\Delta E=-{\frac{\pi }{2\gamma }}\alpha \hbar \omega .
\label{eq06}
\end{equation}
\noindent
It is the MgB$_2$ case indeed on the reason that the mass
ratio is inversely proportional to the hopping integrals ratio:
$M/m\propto t/t_{z}\propto 10$ (see Ref.~\onlinecite{kong}).
For the numerical estimates we will take the experimental value
for the optical phonon frequency $\omega$.
The energy shift $\Delta E$ is negative what means the total lowering of the
band minimum of the hole carriers due to the polaron effect.
Due to $\Delta E$ the energy distance $r$ between the $\sigma$- and
$\pi$-bands is shifted  by a frequency dependent contribution which
is essential for the electronic sector.

\subsection{Electronic Sector}

We start here from the high-temperature
paramagnetic phase of MgB$_2$ system and derive the superconducting
critical temperature from the condition of instability of the normal state
of correlated $\sigma$-electrons with a temperature decrease. Namely,
the $T_c$ is governed by the solution of the
Bethe-Salpeter equation for a vertex $\Gamma_p$
in the Cooper channel in the reference frame of the electron pair:
\begin{equation}
\Gamma _{p}=-T\sum\limits_{n,q}\left[ -2t_{q}+V\left( p-q\right) \right]
G_{\omega _{n}}^{0+}\left( q\right) G_{-\omega _{n}}^{0-}\left( -q\right)
\Gamma _{q},  \label{v1}
\end{equation}
where $G_{\omega _{n}}^{0s}\left( q\right) =1/\left( -i\omega _{n}+\xi
_{q}\right) $ is the normal state Green's function for a strongly correlated
$\sigma$-electron with a spin orientation $s$ and an energy dispersion $\xi
_{q}=ft_{q}-\mu $ with the correlation factor\cite{ivanov2}
\begin{equation}
f=\frac{2w_{1}+w_{2}+3r}{5w_{1}+4w_{2}}.  \label{v2}
\end{equation}
The Matzubara frequencies are given by $\omega _{n}=(2n+1)\pi T$
in Eq.~(\ref{v1}).

The expression for the chemical potential
\begin{equation}
\mu =w_{1}\frac{w_{2}-5r}{5w_{1}+4w_{2}}  \label{v3}
\end{equation}
follows from the equation $n_{\sigma }+n_{\pi }=2$ for the total electron
density per boron at the assumption of a complete ionicity of Mg ions in
Mg$^{2+}\left[{\rm B}^{-}\left(p^{n_p+n_\sigma}\right) \right] _{2}$ system.
In the integral kernel of Eq.~(\ref{v1}) near the $\Gamma$-$A$ line of the
Brillouin zone the Coulomb vertex $V\left( p-q\right) $ can be factorized as
\begin{equation}
V\left( p-q\right) =2\beta t_p t_q,
\label{v4}
\end{equation}
where the parameter $\beta =V/6t^{2}$ labels an effect of the Coulomb
repulsion for the nearest $\sigma$-electrons and the energy dispersion is
$t_p =3t\left[ 1-\left( p_{x}^{2}+p_{y}^{2}\right) /12\right] $
near the $\Gamma$-$A$ line. The kernel of the integral equation (\ref{v1})
does not contain another kinematic vertices\cite{ivanov2}
which could be essential for superconducting condensates at moderate
densities of carriers and/or for specific symmetries of superconducting
order parameters.

Summation over the Matzubara frequencies being performed, Eq.~(\ref{v1})
takes the form:
\begin{equation}
1=\sum\limits_{q}t_{q}\frac{1-\beta t_{q}}{\xi_{q}} \tanh \frac{\xi_{q}}{%
2T_{c}}.  \label{v6}
\end{equation}
The superconducting coupling constant can be written as
\begin{equation}
\lambda =\frac{\mu }{w_{1}f^{2}}\left( 1-\frac{\beta }{f}\mu \right),
\label{lambda}
\end{equation}
so the chemical potential is restricted by the inequality $0\leq \mu \leq
f/\beta$ which generates constraints for the electron structure parameters
$w_{1,2}$ and $r$. The larger is the electron-electron Coulomb repulsion $V \sim
\beta$ the more narrow becomes the superconducting region and the lower
is $T_c$.

Hereafter we will neglect the inter-electron Coulomb repulsion ($\beta=0$).
Then the superconducting critical temperature satisfies
the following equation:
\begin{eqnarray}
1=\int\limits_{\xi(-w_1)}^{\xi(w_1)} d\xi\ \rho(\xi)\frac{\xi+\mu}{\xi f^2}
{\rm tanh}\frac{\xi}{2T_c},
\label{t_c}\end{eqnarray}
where the $\xi (w_{1})$ and $\xi (-w_{1})$ are the energy dispersion values
of $\sigma$-electrons at the top and the bottom of their energy band,
respectively, and $\rho(\xi)$ is the electronic density of states.
Eq. (\ref{t_c}) for the superconducting critical temperature
coincide with one derived in Refs.~\onlinecite{ivanov,ivanov1}
in a different way.

Note that the kinematic superconducting mechanism was also applied
to MgB$_2$ in Ref.~\onlinecite{zaitsev} but with a non-physical negligence
of the $\pi$-electrons role. In our approach the great importance has the
characteristic energy difference $r_{0}$\ between of $\sigma$- and
$\pi$-bands, shifted by $\Delta E$ due to the interaction between
the light holes in the $\sigma$-band and the quasi-$2D$ $E_{2g}$ phonons:
\begin{eqnarray}
r =r_{0}+\Delta E.
\label{v7}
\end{eqnarray}

Due to the chemical bonds nature the hydrostatic pressure
decreases the inter-plane distance more readily than the
in-plane boron-boron distance, so that $dw_{2}/dp \gg dw_{1}/dp$ (c.f.
Ref.~\onlinecite{choi}) and the latter derivative can be neglected in our
estimates. The calculations of the $T_{c}(r)$ demonstrate\cite{ivanov}
that $r\gtrsim -w_{1}/4$ for MgB$_2$. As the next step we differentiate
the integral equation (\ref{t_c}) with respect to
the pressure and take into account Eq.~(\ref{v7}).
Assuming the rectangular density of states $\rho=\theta(w_1^2-\xi^2)/2w_1$
we obtain after the subsequent integration the following expression:
\begin{eqnarray}
\frac{d \ln T_c}{dp} = \frac{1}{5w_1+4w_2} \left[ \frac{2w_1}{T_c} -1
- 5\ln \left( \frac{\gamma_0 w_1}{\pi T_c}\right)\right]  \frac{dr}{dp},
\label{tcp_full}\end{eqnarray}
where $\gamma_0 = \ln C$ with $C=0.577$ being the Euler's constant.
Under the natural assumption $T_c \ll w_1$ one can
keep only the first term in the brackets in Eq.~(\ref{tcp_full}).
Then we arrive at the result:
\begin{equation}
\frac{dT_{c}}{dp}=\frac{2w_1}{5w_{1}+4w_2}\frac{d\Delta E}{dp}.
\label{v8}
\end{equation}

Taking into account that the Fr\"{o}hlich coupling constant
$\alpha \propto 1/\sqrt{\omega}$
we find then from Eq.~(\ref{eq06}):
\begin{equation}
{\frac{d\Delta E}{dp}}=\Delta E{\frac{d\ln (\sqrt{\omega }/\gamma )}{dp}}.
\label{v10}
\end{equation}
The parameter $\gamma =c/2a$ evidently decreases with the pressure. As to the
quasi-$2D$ frequency $\omega $ we use the estimate of a quasi-harmonic
phonon via the mode Gr\"{u}neisen parameter $d\ln \omega /dp=G/B_{0}$. The
quantity $B_{0}$ is said to be the bulk modulus. It follows then that both
$\omega $ and $\sqrt{\omega }/\gamma $ increase with the pressure. These
findings are compatible with an experimental study\cite{goncharov,meletov}
of Raman spectra and lattice parameters of MgB$_{2}$ under pressure at
room temperature and with theoretical estimations\cite{kobayashi} of the
strong influence of the pressure on $E_{2g}$ mode. Finally we conclude that
$d\Delta E/dp<0$ and $dT_{c}/dp<0$.

In an explicit form the baric derivative for the superconducting critical
temperature can be written as follows:
\begin{eqnarray}
\frac{d T_{c}}{dp} = - \frac{|\Delta E|}{5+4w_2/w_1}
\left( -\frac{d\ln \gamma^2 }{dp}+\frac{G}{B_{0}}\right).
\label{v11}
\end{eqnarray}

\section{Numerical Results and Conclusions}

For numerical calculations we estimate the polaron
anisotropy parameter as $\gamma = c/2a = 0.57$.
The known value of its derivative\cite{jorgensen} is
$d\ln \gamma ^{2}/dp=2d\ln \left( c/2a\right)
/dp\simeq -2.4\cdot 10^{-3}\ GPa^{-1}$.
The bulk modulus is measured\cite{li} to be
$B \simeq 114\ GPa$. The Gr\"{u}neisen parameter
reported in Ref.~\onlinecite{goncharov} equals $G=2.9 \pm 0.3$
for the measured Raman active
$E_{2g}$-phonon mode with the energy $\hbar\omega=76.7~meV$
which we will use in Eq.~(\ref{v11}).
In particular, the polaron self-energy $\Delta E = -211.4\alpha~meV$.
The realistic value for the MgB$_{2}$ energy
bandwidths ratio of $\pi$- and $\sigma$-electrons
is $w_2/w_1 \simeq 18/9=2$ (see Ref.~\onlinecite{antropov}).

Putting these magnitudes in Eq.~(\ref{v11})
we obtain the estimate for the derivative of $T_{c\text{ }}$ with respect
to pressure:
\begin{equation}
\frac{dT_{c\text{ }}}{dp} \simeq -\alpha (5.2\pm 0.9)\frac{K}{GPa}.
\label{v12}
\end{equation}
The uncertainty of the result comes from the experimental deviation errors
in the Gr\"{u}neisen parameter. The comparison with the low pressure
result\cite{lorenz,tissen}
$dT_{c}/dp = -1.2~K/GPa$ leads at the interval $\alpha = 0.20 - 0.28$
for the Fr\"{o}hlich electron-phonon coupling constant. The comparison
with the result\cite{lorenz,prassides,saito}
$-2~K/GPa$ leads subsequently at the estimate $\alpha = 0.33 - 0.47$.

\begin{figure}[t]
  \hspace*{-1.0cm}
\epsfysize=2.7in
\epsfbox{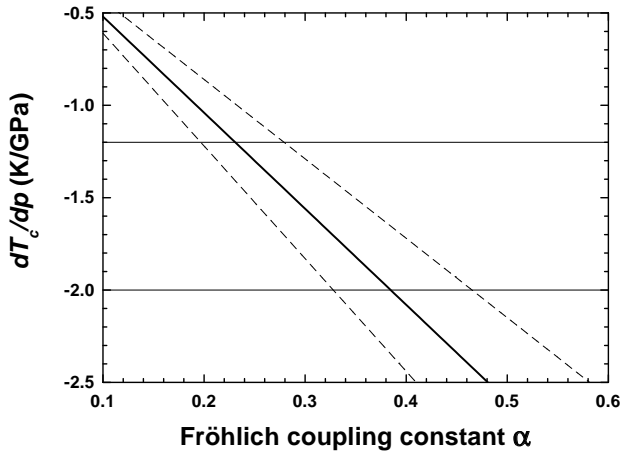}
\caption{Dependence of the baric derivative of the superconducting
critical temperature on the Fr{\protect\"o}hlich electron-phonon coupling
constant (solid line). The dashed lines show error bars for $\alpha$
due to the uncertainty in the Gr{\protect\"u}neisen
parameter\protect\cite{goncharov}.
The thin solid lines show experimental
values\protect\cite{lorenz,tissen,prassides,saito} for $dT_c/dp$. }
\label{fig1}
\end{figure}

These results are shown in Fig.~\ref{fig1}. The solid line
presents the dependence of the baric derivative $dT_c/dp$
on the electron-phonon coupling constant $\alpha$.
The thin solid lines set the experimentally
measured values for $dT_c/dp$ and the dashed lines
show error bars for $\alpha$ at a taken value of the baric derivative.

We have to note that the numerical results can be changed if one
scales the frequency shift of the in-plane mode with the variation
of the interatomic bond distance or lattice parameter\cite{hanfland}.
Then the Gr\"{u}neisen parameter takes even the larger
value\cite{goncharov}
$G=3.9\pm 0.4$ and our Eq.~(\ref{v12}) takes the form:
\begin{eqnarray}
\frac{dT_c}{dp} \simeq -\alpha (6.9\pm 1.1)\frac{K}{GPa}.
\label{v13}\end{eqnarray}
Subsequently, the estimates for the electron-phonon coupling constant
will be different as well: $\alpha = 0.15 - 0.21$ for
$dT_{c}/dp = -1.2~K/GPa$ and $\alpha = 0.25 - 0.34$ for
$dT_{c}/dp = -2~K/GPa$.

Combining the numerical results for $\alpha$
we may conclude that the electron-phonon coupling constant
in MgB$_2$ is in the range $\alpha = 0.15 - 0.45$ which seems
to be quite reasonable interval for the assumed weak-coupling regime of
a large hole polaron.

To resume, we proposed a model in which large anisotropic polarons
play an important role decreasing the energy distance $r$ between
the $\sigma$- and $\pi$-bands. The polaron anisotropy is governed by the
introduced parameter $\gamma$ which depends on the
geometry of the MgB$_2$ system influenced by the pressure.
Besides, the phonon frequency increases with the pressure which results
also in a decrease of the polaron radius. Thus, the superconducting
properties of the MgB$_2$ system are influenced both by the geometry of
the crystal and by the polaron well depth and size.

The superconducting instability is driven
by the non-phonon kinematic mechanism with $T_c$ depending
on the energy difference $r$ between the $\sigma$- and $\pi$-bands.
The quantity $r$ incorporates all electron-phonon effects in our model.
As the result the baric derivative $dT_c/dp$ is calculated as a function
of the Fr\"{o}hlich coupling constant $\alpha$. At realistic values
of $\alpha$ our calculations agree with the experiments.
It follows from the arguments discussed in the present paper that
the pressure measurements
provide us with valuable tests to establish the acceptable model
for MgB$_2$, to understand the mechanism
of the superconductivity in this material and to estimate its
Fr\"{o}hlich electron-phonon coupling constant.

\acknowledgments
We are grateful to K.~Kikoin for stimulating discussions.
M.A.S. thanks Universiteit Antwerpen --- TFVS for the hospitality during
his visit to Belgium. The work is supported by GOA BOF UA 2000, IUAP and
FWO-V project No. G.0274.01.

\end{document}